\documentclass[english]{article}

\usepackage{amsmath}
\usepackage{amssymb}
\usepackage{cite}
\usepackage{graphicx}
\usepackage{color}

\def\equ#1{(\ref{#1})}
\definecolor{Red}{rgb}{1.0,0.0,0.0}

\def\be#1{\begin{equation}\label{#1}}
\def\ee{\end{equation}}
\makeatletter

\usepackage{babel}

\begin{document}
\begin{titlepage}
\thispagestyle{empty}

\bigskip

\begin{center}
\noindent{\Large \textbf
{New Analytical Solutions for Bosonic Field Trapping in Thick Branes}}\\

\vspace{0,5cm}

\noindent{R. R. Landim ${}^{a}$, G. Alencar ${}^{a}$\footnote{e-mail: geovamaciel@gmail.com }, M. O. Tahim ${}^{b}$ and R.N. Costa Filho ${}^{a}$}

\vspace{0,5cm}
 {\it ${}^a$Departamento de F\'{\i}sica, Universidade Federal do Cear\'{a}-
Caixa Postal 6030, Campus do Pici, 60455-760, Fortaleza, Cear\'{a}, Brazil.

\vspace{0.2cm}
 }
{\it ${}^b$Universidade Estadual do Cear\'a, Faculdade de Educa\c c\~ao, Ci\^encias e Letras do Sert\~ao Central- 
R. Epit\'acio Pessoa, 2554, 63.900-000  Quixad\'{a}, Cear\'{a},  Brazil.
}
\end{center}

\vspace{0.3cm}

\begin{abstract}
New analytical solutions for gravity, scalar and vector field localization in Randall-Sundrum(RS) models are found. A smooth version of the warp factor with an associated function $f(z)=\exp(3A(z)/2)$  inside the walls ($|z|<d$) is defined, leading to an associated equation and physical constraints on the continuity and smoothness of the background resulting in a new space of analytical solutions. We solve this associated equation analytically for the parabolic and P\"oschl-Teller potentials and analyze the spectrum of resonances  for these fields. By using the boundary conditions we are able to show that, for any of these solutions, the density probability for finding a massive mode in the membrane has a universal  behavior for small values of mass given by $|\psi_m(0)|^2=\beta_1m+\beta_3m^3+\beta_L m^3\log(m)+\cdots$. As a consequence, the form of the leading order correction, for example, to the Newton's law is general and does not depend on the potential used. At the end we also discuss why complications arises when we try to use the method to find analytical solutions to the fermion case.
\end{abstract}
\end{titlepage}

\section{Introduction}

After the seminal work of Randall and Sundrum(RS)\cite{Randall:1999vf} several other results have been developed based on the idea of membranes as topological defects and its implications for brane world physics\cite{Bazeia:2005hu,Liu:2009ve,Zhao:2009ja,Liang:2009zzf,Zhao:2010mk,Zhao:2011hg}. In these models one must determine the space of solutions of a Schr\"odinger equation with a specific
potential which depends on the warp factor.  That is, one needs to solve a Sturm-Liouville problem to find eigensolutions with eigenvalues. A particular application of this  kind of model is in the study of gravity trapping in a finite thickness domain wall \cite{Cvetic:2008gu}, where a constant potential in the region near/over the membrane is chosen in order to find analytical solutions.  The benefits of such analytical solutions are worth because it allows explicitly analysis of the Kaluza-Klein masses and opens up possibilities for analytical studies of fermionic modes. These analysis and possibilities can be extended even more if different potentials for the graviton wave function modes can be solved analytically. 

In this manuscript, we present a new explicit integrable Schr\"odinger potentials for the graviton wave function modes parameterized by the thickness of the wall. The warp factor is chosen in order to be continuous in the boundary of two regions of the space-time. These two regions basically describe the interaction right near/over the membrane location and interactions far from the membrane. We find an equation that drives the profile of the brane. With this we show that  the function used in \cite{Cvetic:2008gu} is just a particular solution of the equation presented here. This reveals a new space of analytical solutions and, as direct consequences, new zero modes, Kaluza-Klein modes, new resonance behavior, and so on. The new analytical solutions are encoded in a Schr\"odinger-like differential equation with zero eigenvalue. With this is possible to show that, for small values of $m$, the  probability density for finding 
a mass mode in the membrane does not depend on the chosen potential. In this way, the leading order correction of the four dimensional Newton's law, for example, has  a general expression that does not depend on the potential used. In the following lines we discuss how to apply this method to study the physics of gravitational fields, scalar fields and gauge vector fields in the RS scenario 

\section{The associated equation}

To start our reasoning we remember the fact that the mass spectrum of the gravity field is driven by a
Schr\"odinger like equation \cite{Landim:2011ki}
\begin{equation}\label{Sch}
-\psi''_m(z)+U(z)\psi_m(z)=m^2\psi(z),
\end{equation}
where the effective potential depends on the warp factor, $A(z)$, as below
\begin{equation}\label{U(A)}
U(z)=\frac{3}{2}A''(z)+\frac{9}{4}A'(z)^{2},
\end{equation}
and the metric in the conformal coordinate is given by $ds^2=e^{2A(z)}(dx^2+dz^2)$. By analyzing the above equation it has been shown that the zero mode $(m=0)$ is trapped in the membrane. However as $\lim_{z\to\infty}U(z)=0$, the massive modes are not localized. An important phenomenological aspect related to this is the appearance of resonances. This allows for the possibility of unstable massive modes that could be seen in the membrane. Most of these studies have been performed numerically by considering smooth versions of the RS model. These smooth versions can be obtained, for instance, by considering the membrane as a topological defect generated by a scalar field. In this scenario the condition imposed is that for large $z$ the RS warp factor is recovered. These models have many interesting properties and have been widely studied over the last decade\cite{Bazeia:2005hu,Liu:2009ve,Zhao:2009ja,Liang:2009zzf,Zhao:2010mk,Zhao:2011hg}.

Another way to get a smooth version of the RS model is by considering the brane as a thick domain wall \cite{Cvetic:2008gu,Alencar:2012en}. In these papers the potential used depends on one parameter $0\leq
x\leq\pi/2$ which determines the thickness of the membrane.  In order to get the desired smooth version, $A$ and $A'$ must be continuous and this imposes some restrictions on the form of $A(z)$ in the membrane.
The choice of \cite{Cvetic:2008gu} was $A(z)=\frac{2}{3}\ln cos(\sqrt{V_{0}}|z|)$ and this give the effective potential $U=-V_{0}$. With this, the resonances of the model were studied analytically in detail.

In order to obtain a wider class of new exact solutions, we divide the warp factor $A(z)$ in two regions, $|z|\leq d$ and $|z|\geq d$:
\begin{eqnarray}\label{A}
A(z)=
\begin{cases}
\frac{2}{3}\ln[f_0(z)], &|z|\leq d,\\
~\\
\ln\left(\frac{1}{|z|+\beta}\right), & |z|\geq d,
\end{cases}
\end{eqnarray}
where we defined $f_0(z)$ as the associated function.  The analytical solution for $|z|\geq d$ is 
already known \cite{Cvetic:2008gu,Alencar:2012en}, then we focus in analytical solutions for $|z|\leq d$. 
With Eqs. (\ref{A}) and (\ref{U(A)}) we get that $f_0(z)$  satisfies 
the associated equation 
\begin{equation}\label{f}
-f''_0(z)+U(z)f_0(z)=0,
\end{equation}
this is exactly the effective Schr\"odinger equation (\ref{Sch})
for $m^2=0$. In order to implement the boundary conditions, we restrict to even
functions $f_0(z)=f_0(-z)=g_0(z)$ with $U(z)=U(-z)$ in \equ{f}, guarantying that
the boundary conditions are satisfied in both edges of the brane,
$z=d$ and $z=-d$.  The condition $A(0)=0$ implies that
$g_0(0)=1$. Since $g_0(z)$ is an even function we also have $A'(0)=0$. This conditions fix  completely the solution of \equ{f}.   With the above considerations and by imposing continuity of $A(z)$ and $A'(z)$ at $z=\pm d$ we obtain
\begin{align}\label{cont}
\frac{2}{3}g'_0(d)&=-g_0(d)^{1+2/3},\\ 
\beta&=-d+\frac{1}{g_0(d)^{2/3}},
\end{align}
with the conditions $g'_0(d)<0$  and $g_0(z)$ positive and limited in $|z|\leq d$. In the last section it will become clear why we have written $5/3=1+2/3$.

\section{Analytical Solutions}

As a first example, let us consider a constant potential $-|V_0|$ in the region $|z|\leq d$ \cite{Cvetic:2008gu}. We know that the
solution for the Sch\"odinger equation (\ref{Sch}) is a linear
combination of $\cos(\sqrt{m^2+|V_0|}z)$ and
$\sin(\sqrt{m^2+|V_0|}z)$. The even
solution for $m=0$ that satisfies $g_0(0)=1$ is $\cos(\sqrt{|V_0|} z)$. The
conditions $g'_0(d)<0$ and $g_0(z)>0$ implies that
$0<\sqrt{|V_0|}d<\pi/2$. This is the solution introduced in
\cite{Cvetic:2008gu,Alencar:2012en}.
As a straightforward application of the method developed here lets examine the harmonic oscillator with Schr\"odinger equation
\begin{equation}\label{pot-osc}
-\psi''_m(z)+z^2\psi_m(z)=m^2\psi_m(z).
\end{equation}
The above equation can be cast in the form of a Kummer equation
\cite{AbramowitzStegun64}, by writing $\psi_m(z)=e^{-z^2/2}w_m(z)$ and
next using the transformation $u=z^2$, we obtain
\begin{equation}
uF''(u)+(b-u)F'(u)-aF(u)=0,
\end{equation}
with $b=1/2$ and $a=(1-m^2)/4$. Then, solutions of \equ{pot-osc} are
linear combinations of $g_m(z)=e^{-z^2/2}F_1(a;\frac{1}{2};z^2)$ and
$h_m(z)=ze^{-z^2/2}F_1(a+\frac{1}{2};\frac{3}{2};z^2)$, where
$F_1(a;b;z)$ is the Kummer confluent hypergeometric function. From now on we will use for the even(odd) solution in $|z|\leq d$ the notation $g_m(z)$($h_m(z)$).  The even solution for $m=0$ with $g_0(0)=1$ is
$g_0(z)=e^{-z^2/2}F_1(\frac{1}{4};\frac{1}{2};z^2).$

In fact, we can find a large new class of solutions simply considering $U(z)=az^2+b$, with $a>0$. Using the above-mentioned steps we find the pair of solutions
\begin{align}
&g_m(z)=e^{-\sqrt{a}z^2/2}F_1\left(\frac{b}{4\sqrt{a}}-\frac{m^2}{4\sqrt{a}}+\frac{1}{4};\frac{1}{2};{\sqrt{a}z^2}\right),
\\
&h_m(z)=ze^{-\sqrt{a}z^2/2}F_1\left(\frac{b}{4\sqrt{a}}-\frac{m^2}{4\sqrt{a}}+\frac{3}{4};\frac{3}{2};{\sqrt{a}z^2}\right),
\end{align}
and $W(g,h)(z)=1$, where $W(f_1,f_2)(x)=f_1(x)f_2'(x)-f_1'(x)f_2(x)$ is the Wronskian of $f_1,f_2$. It is worthwhile to mention that the Wronskian is constant for Schr\"odinger-like equations. The even  solution satisfies $g_0(0)=1$ and the value of the constants $a,b$ and $d$ are related in order to give $g'_0(d)<0$ and $g_0(z)>0$.  As an example, for $b=-5$ and $a=1$, $g_0(z)$ is 
positive defined for $|z|<0.707107$ and $d=0.243928$. 

The method described here can be applied to many other cases known in physics. One possibility 
is the problem for a particle in a box subject to a constant field. This is described by a linear potential$U(z)=az$ giving rise to the Airy functions with solutions $Ai(z)$ and $Bi(z)$. However, this potential do not satisfies the condition of being even. Another class of analytical solutions can be found by considering exponential potentials. From these the only even one is the P\"oschl-Teller potential, 
where the Schr\"odinger-like equation  is 
\begin{equation}\label{pt}
 \psi_m''(z)+(m^2+a^2b(b+1)\mbox{sech}^2(a z))\psi_m(z)=0.
\end{equation}
Rewriting $\psi(z)=w(z)/\cosh^b(a z)$ and next using the transformation $u=-\sinh^2(a z)$, we can write \equ{pt} as a hypergeometric differential equation
\begin{equation}\nonumber
u(1-u)F''(u)+(\gamma-(\alpha+\beta+1)u)F'(u)-\alpha\beta F(u)=0,
\end{equation}
where $\gamma=1/2$, $\alpha=(-b+im/a)/2$ and $\beta=-(b+im/a)/2$. Therefore the linearly independent solutions of \equ{pt} are $g_m(z)=F(\alpha,\beta;1/2;-\sinh^2(a z))/\cosh^b(a z)$ and
$h_m(z)=\sinh(a z)F(\alpha+1/2,\beta+1/2;3/2;-\sinh^2(a z))/\cosh^b(a z)$ with $W(g_m,h_m)(z)=a$ and $g_0(0)=g_m(0)=1$. 

\begin{figure}[h!]
\centering
\includegraphics[scale =0.5]{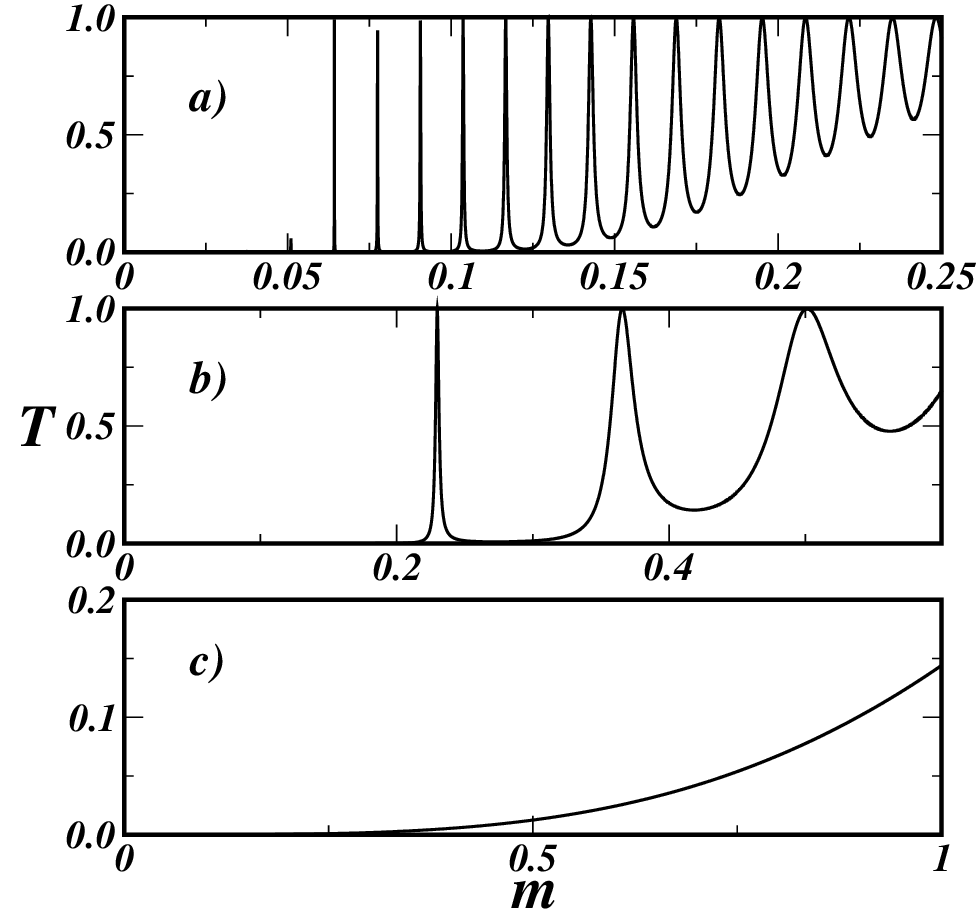}
\caption{The transmission coefficient for the P\"oschl-Teller potential with  a) a=0.01,b=1 (d=115.369), b) a=0.1,b=1 (d=10.1721), and c) a=1,b=1 (d=0.537862).}
    \label{pteller}
\end{figure}

After fixing the background with the above method we turn our attention to the gravity resonances. The interesting fact about this background is that we automatically have exact solutions to equation \equ{Sch} just by not fixing $m=0$. With this we get an analytical expression for our resonances. As we are interested in resonances we must consider a plane wave coming from $-\infty$. This plane wave will collide with the membrane and will generate a reflected and a transmitted wave. Therefore, for $z<-d$ we must have a linear combination of waves moving to the left and to the right. For $z>d$ we must have only one wave moving to the right. In order to analyze the resonances we fix the coefficient of the incoming
wave equal to one. In this way, the Schr\"odinger-like equation \equ{Sch} has the solution
\begin{eqnarray}\label{psim}
\psi_m(z)=
\begin{cases}
A_mg_m(z)+B_mh_m(z), &|z|\leq d,\\
 F_m(z)+C_mE_m(z), & z\leq -d,\\
D_m F_m(z), & z\geq d  
\end{cases}
\end{eqnarray}
where $E_m(z)=\sqrt{\frac{\pi u}{2}}H_{2}^{(2)}(u),F_m(z)=\sqrt{\frac{\pi u}{2}}H_{2}^{(1)}(u)$, with $u=m(|z|+\beta)$. We shall use the convention 
that $g_m(0)=1$. Taking the continuity of the wave function and its derivative at $z=\pm d$
%
and using the fact that $g_m(z),E_m(z),F_m(z)$ are even and $h_m(z)$ are odd we finally obtain the transmission coefficient
\begin{figure}[h!]
\centering
\includegraphics[scale =0.5]{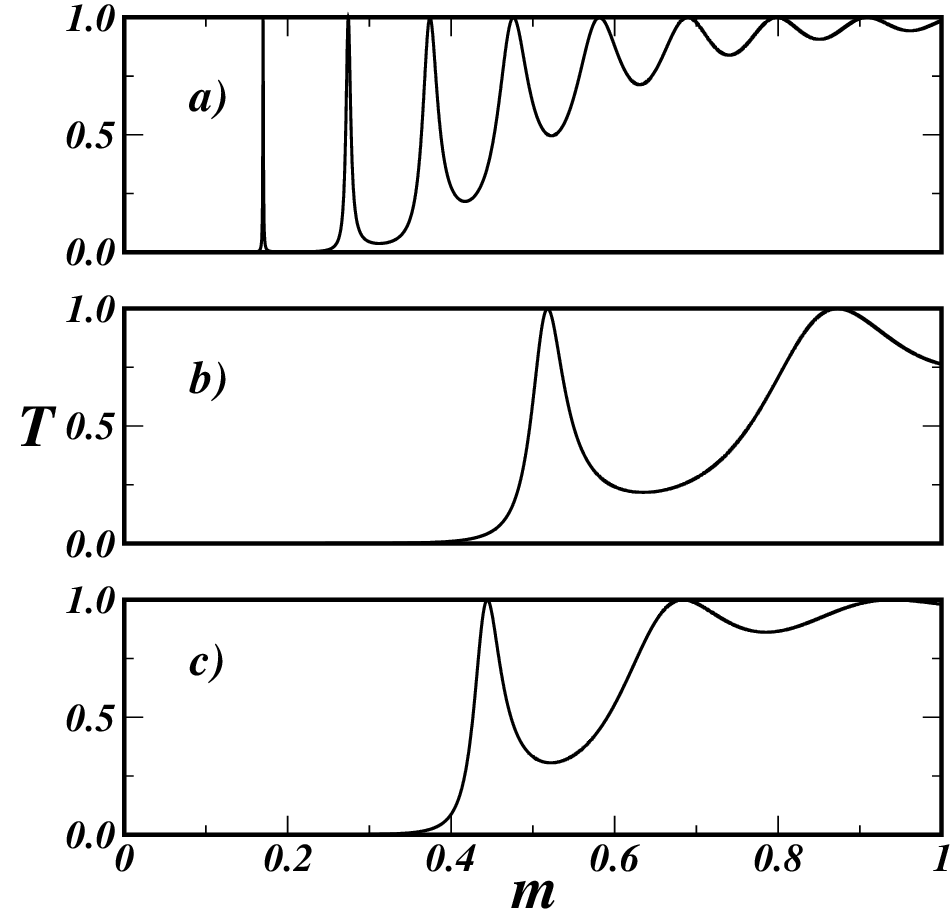}
\caption{The transmission coefficient for parabolic potential with: a) a=0.001,b=-0.1 (d=3.8122) ; b) a=0.01,b=-0.1 (d=5.49082) and c) a=0.00001,b=-0.01 (d=13.9833).}
\label{parabolic}
\end{figure}
\begin{equation}\label{analictt}
T(m)=\frac{m^2|W(g_m,h_m)(d)|^2}{|W(F_m,g_m)(d)W(F_m,h_m)(d)|^2}.
\end{equation}
With this expression and the solutions found before we can easily obtain graphics for the transmission coefficients. In Fig. \ref{parabolic} we show the trasmission for the parabolic potential. As expected, depending on the parameters we can have different peaks of resonances. It is important to point that as narrow the resonance peak  the stronger is the signal of the mass mode. In Fig. \ref{parabolic} we see that happening for the first peaks. For the P\"oschl-Teller potential Fig. \ref{pteller} shows resonances depending on the parameters. For the parameters used the resonances are more peaked, what is phenomenologically more interesting. It is clear that the potentials present resonances with different characteristics.

\section{Correction to the Newton's Law}

\begin{figure}[h!]
\centering
\includegraphics[scale =0.5]{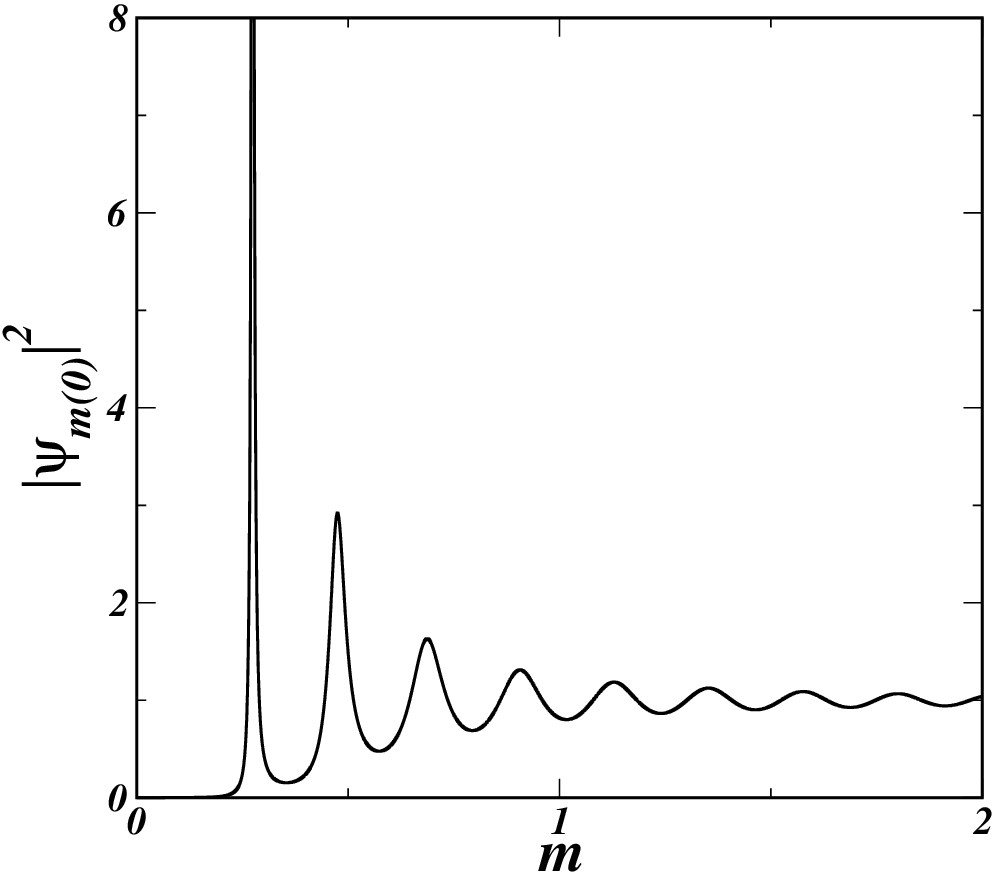}
\caption{The amplitude probability as  function of $m$.}
    \label{psi}
\end{figure}

In order to analyze the deviation of Newton's law, we must calculate the probability density of
the wave function at $z=0$:
\begin{equation}
|\psi_m(0)|^2=|A_m|^2=\frac{m^2}{|W(F_m,g_m)(d)|^2}.
\label{psi0}
\end{equation}
The above expression is plotted in Fig. \ref{psi} for the parabolic potential with $a=0.00001$ and $b=-0.01$.

For small values of $m$, $F_m(z)$  has the following form for $z>0$:
\begin{eqnarray}
F_m(z)\approx a_{-3/2}(z)m^{-3/2}+a_{1/2}(z)m^{1/2}+\\ \nonumber
a_{5/2}(z)m^{5/2}+a_{L}(z)m^{5/2}\log(m),
\end{eqnarray}
where 
\begin{align}
 a_{-3/2}(z)&=-2i\sqrt{\frac{2}{\pi}}(z+\beta)^{-3/2},\\
 a_{1/2}(z)&=-i\displaystyle{\frac{(z+\beta)^{1/2}}{\sqrt{2\pi}}},\\
 a_{5/2}(z)&=\displaystyle{(\gamma_0+4i\log(z+\beta))\frac{(z+\beta)^{5/2}}{16\sqrt{2\pi}}},\\
 a_L(z)&=4i\displaystyle{\frac{(z+\beta)^{5/2}}{16\sqrt{2\pi}}},
\end{align}
with $\gamma_0=2\pi-3i+4i\gamma-4i\log(2)$ and $\gamma$ is the Euler-Mascheroni constant. The function $g_m(z)$ is well 
defined for $m=0$, i.e, $g_m(z)$ do not have poles at $m=0$. Therefore, for small $m$ we can expand $g_m(z)$ in a power series 
in $m^2$: $g_m(z)=g_0(z)+b_1(z)m^2+b_2(z)m^4+\cdots$. The wronskian $W(a_{-3/2},g_0)(d)$ is zero due to the boundary conditions \equ{cont}. Then we have $W(F_m,g_m)(d)=\sqrt{m}\left(\alpha_0+\alpha_2 m^2+\alpha_L m^2\log(m)+\cdots\right)$. With this we arrive to the following expansion for $m<<1$:
\be{Aexp} 
\left|\psi_m(0)\right|^2=\beta_1m+\beta_3m^3+\beta_L m^3\log(m)+\cdots,
\ee
where $\beta_1, \beta_3,\beta_L$ are coefficients that depend on the potential used through $a_i(d),b_i(d)$. For large $m$, $F_m(z)\sim e^{imz}$ and $g_m(z)\sim \cos mz$ consequently $|A_m|^2\rightarrow 1$ when $m\rightarrow\infty$.

To calculate the deviation of Newton's law we have to plug the expansion (\ref{Aexp}) in the expression 
\be{ndev}
\delta(r)=\int_0^\infty |\psi_m(0)|^2 e^{-mr}dm,
\ee
and integrate from $(0,M_c)$ for the small $M$ expansion and $(M_c,\infty)$ with $M_c<1$. The integration gives us 
\begin{eqnarray}
 \delta(r)=~&&\frac{\beta_1}{r^2}\left({1-e^{-M_c r} (1+M_c r)}\right)\\ \nonumber
 &&+\frac{\beta_3}{r^4} \left(6-e^{-M_c r}((M_cr)^3+3(M_cr)^2+6M_cr +6)\right) \\ \nonumber
 && \frac{\beta_L}{r^4}\left(11-6\gamma+6\log(M_c)-\Gamma(4,M_cr)\log(M_c)\right.\\ \nonumber
 &&-\left. 6\log(M_cr)-G^{3,0}_{2,0}({1,1},{0,0,4}|M_cr)\right)+\frac{e^{-M_cr}}{r}, 
\end{eqnarray}
where $\Gamma(y,x)$ is the complete gamma function and $G^{3,0}_{2,0}({1,1},{0,0,4}|x)$ is the Meijer G-function.

\section{Scalar, Vector and Fermion Fields}

\begin{figure}[h!]
\centering
\includegraphics[scale =0.5]{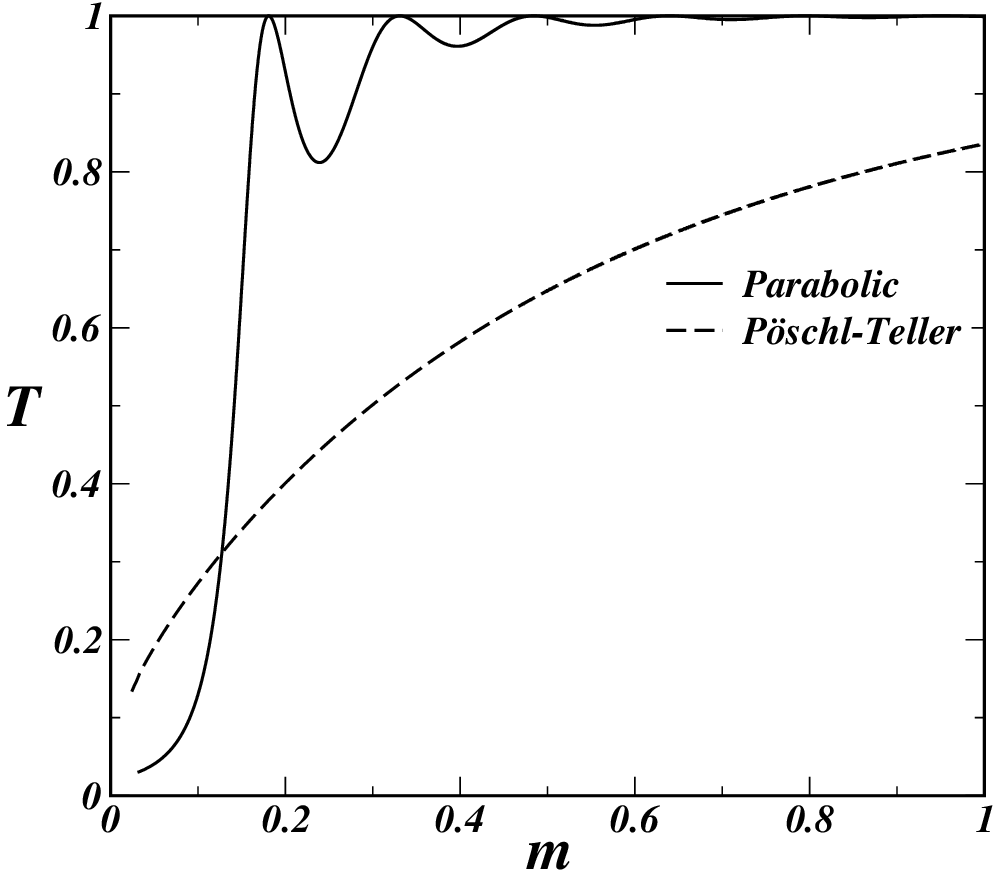}
\caption{The transmission coefficient for the parabolic ( d=9.9967, a = 0.00001; b = -0.01) and P\"oschl-Teller potentials (d=0.222377,a=1,b=1).}
    \label{vetor}
\end{figure}

In a previous work the present authors have found analytical solutions for other bosonic fields in the potential well case $U=-V_0$ \cite{Alencar:2012en}. We can use this and the above method to find new analytical solutions to the scalar and vector fields. What was pointed out  in that work is that for any effective potential which takes the form $U(z)=cA''+c^2A'^2$, an analytical solution can be found easily just by changing some parameters of the theory. We have also shown that for the scalar and vector fields $c=3/2$ and $c=1/2$ respectively. We also pointed that the parameter for the scalar field is the same of the gravity field, therefore all results are identical and we just need to consider the vector case here. The only thing we need to do is to change $2/3\to 1/c$ in Eqs. (\ref{A}) and (\ref{cont}). The effect of this for the exterior solution is to change the order of the Hankel function $2\to 1/2+c$. The interior region is changed throughout the contour conditions (\ref{cont}). 
Therefore all the main results, namely  Eqs. (\ref{psim}), (\ref{analictt}), (\ref{psi0}) and (\ref{Aexp}) are kept unchanged up to the above change in the parameters. Therefore, the expansion for small masses have the same general behavior also for the scalar and vector fields. In Fig. \ref{vetor} we show the transmission coefficients for the vector case. Note that the change of the potential can give very different phenomenological results for a observer in the brane.  

 At this point some discussion about the fermion case is important. The main difference between this and the bosonic case is the fact that the way to get a nontrivial potential is through the addition of a coupling with a scalar field. This is described in very well in Ref. \cite{Li:2010dy}, where it is used the constant potential of \cite{Cvetic:2008gu}. The fact is that when this coupling is considered two additional complications arise. One is the fact that the following relation $\phi'(z)^2=3(A'(z)^2-A''(z))$ must be satisfied \cite{Li:2010dy}. Therefore we must impose the continuity of the second derivative of the warp factor if we wants a continuous first derivative of the scalar field. Another and more difficult problem is that the effective potential is given by
\begin{equation}\label{U(A)}
U_{\pm}(z)=(\eta F(\phi(z))e^{A(z)})^2{\pm}\frac{d}{dz}(\eta F(\phi(z))e^{A(z)}),
\end{equation}
which do not posses a general analytical solution even for the simplest case of the constant potential\cite{Li:2010dy}.

\section{Summary and Conclusions}
In summary we have described a prescription to find new analytical solutions in thick brane models. Two cases are presented as examples: we have shown results for the parabolic and the P\"oschl-Teller potentials. We have studied the problem of resonances in co-dimension one brane world, finding the mass spectra of the new profiles for the thick branes proposed. For these potentials, the transmission coefficients for the masses are plotted expressing the different peculiarities of each potential. It is important to stress that this work is a generalization of the article \cite{Cvetic:2008gu}. We have shown that the solution used by Mirjam et.al.  is a particular solution of a wider class described here, including the cases of scalar and vector fields. We also have shown that the correction in the Newton`s law found by them is valid for any of these wider class of solutions. The results found here opens up the possibility to explore different backgrounds to study gravity, scalar and vector field resonances in the Randall-Sundrum model. For the fermionic case it is not possible  to aplly the same methods. Only numerical results were found and we hope, in next studies, to go one step further in finding pure analytical solutions.

\section*{Acknowledgments}

We acknowledge the financial support provided by Funda\c c\~ao Cearense de Apoio ao Desenvolvimento Cient\'\i fico e Tecnol\'ogico (FUNCAP), the Conselho Nacional de 
Desenvolvimento Cient\'\i fico e Tecnol\'ogico (CNPq) and FUNCAP/CNPq/PRONEX.

\providecommand{\href}[2]{#2}\begingroup\raggedright\endgroup


\begin{thebibliography}{10}

\bibitem{Randall:1999vf}
Lisa Randall and Raman Sundrum,
  \href{http://dx.doi.org/10.1103/PhysRevLett.83.4690}{{\em Phys.Rev.Lett.}
  {\bfseries 83} (1999) 4690--4693},
\href{http://arxiv.org/abs/hep-th/9906064}{{\ttfamily arXiv:hep-th/9906064
  [hep-th]}}.

\bibitem{Bazeia:2005hu}
D.~Bazeia and L.~Losano,
  \href{http://dx.doi.org/10.1103/PhysRevD.73.025016}{{\em Phys.Rev.}
  {\bfseries D73} (2006) 025016},
\href{http://arxiv.org/abs/hep-th/0511193}{{\ttfamily arXiv:hep-th/0511193
  [hep-th]}}.

\bibitem{Liu:2009ve}
Yu-Xiao Liu, Jie Yang, Zhen-Hua Zhao, Chun-E Fu, and Yi-Shi Duan,
  \href{http://dx.doi.org/10.1103/PhysRevD.80.065019}{{\em Phys.Rev.}
  {\bfseries D80} (2009) 065019},
\href{http://arxiv.org/abs/0904.1785}{{\ttfamily arXiv:0904.1785 [hep-th]}}.

\bibitem{Zhao:2009ja}
Zhen-Hua Zhao, Yu-Xiao Liu, and Hai-Tao Li,
  \href{http://dx.doi.org/10.1088/0264-9381/27/18/185001}{{\em
  Class.Quant.Grav.} {\bfseries 27} (2010) 185001},
\href{http://arxiv.org/abs/0911.2572}{{\ttfamily arXiv:0911.2572 [hep-th]}}.

\bibitem{Liang:2009zzf}
Jun Liang and Yi-Shi Duan,
\href{http://dx.doi.org/10.1016/j.physletb.2009.10.012}{{\em Phys.Lett.}
  {\bfseries B681} (2009) 172--178}.

\bibitem{Zhao:2010mk}
Zhen-Hua Zhao, Yu-Xiao Liu, Hai-Tao Li, and Yong-Qiang Wang,
  \href{http://dx.doi.org/10.1103/PhysRevD.82.084030}{{\em Phys.Rev.}
  {\bfseries D82} (2010) 084030},
\href{http://arxiv.org/abs/1004.2181}{{\ttfamily arXiv:1004.2181 [hep-th]}}.

\bibitem{Zhao:2011hg}
Zhen-Hua Zhao, Yu-Xiao Liu, Yong-Qiang Wang, and Hai-Tao Li,
  \href{http://dx.doi.org/10.1007/JHEP06(2011)045}{{\em JHEP} {\bfseries 1106}
  (2011) 045},
\href{http://arxiv.org/abs/1102.4894}{{\ttfamily arXiv:1102.4894 [hep-th]}}.

\bibitem{Cvetic:2008gu}
Mirjam Cvetic and Marko Robnik,
  \href{http://dx.doi.org/10.1103/PhysRevD.77.124003}{{\em Phys.Rev.}
  {\bfseries D77} (2008) 124003},
\href{http://arxiv.org/abs/0801.0801}{{\ttfamily arXiv:0801.0801 [hep-th]}}.

\bibitem{Landim:2011ki}
R.R. Landim, G.~Alencar, M.O. Tahim, and R.N. Costa~Filho,
  \href{http://dx.doi.org/10.1007/JHEP08(2011)071}{{\em JHEP} {\bfseries 1108}
  (2011) 071},
\href{http://arxiv.org/abs/1105.5573}{{\ttfamily arXiv:1105.5573 [hep-th]}}.

\bibitem{Alencar:2012en}
G.~Alencar, R.R. Landim, M.O. Tahim, and R.N.~Costa Filho,
  \href{http://dx.doi.org/10.1007/JHEP01(2013)050}{{\em JHEP} {\bfseries 1301}
  (2013) 050},
\href{http://arxiv.org/abs/1207.3054}{{\ttfamily arXiv:1207.3054 [hep-th]}}.

\bibitem{AbramowitzStegun64}
Milton Abramowitz and Irene~A. Stegun, {\em Handbook of Mathematical Functions
  with Formulas, Graphs, and Mathematical Tables}.
\newblock Dover Publications, New York, 1964.

\bibitem{Li:2010dy} 
  H.~-T.~Li, Y.~-X.~Liu, Z.~-H.~Zhao and H.~Guo,
  Phys.\ Rev.\ D {\bf 83}, 045006 (2011)
  [arXiv:1006.4240 [hep-th]].
\end{thebibliography}
\end{document}